\documentclass[10pt, conference, compsocconf]{IEEEtran}
\ifCLASSINFOpdf
   \usepackage[pdftex]{graphicx}
   
\usepackage{times}
\usepackage{graphicx}
\usepackage{amsmath}
\usepackage{amssymb}
\usepackage{multirow}
\usepackage{verbatim}
\usepackage{array}
\else
\fi
\begin{document}
%
\title{A Fast, Semi-Automatic Brain Structure Segmentation Algorithm for Magnetic Resonance Imaging}


\author{\IEEEauthorblockN{Kevin Karsch, Qing He, Ye Duan}
\IEEEauthorblockA{Department of Computer Science\\
University of Missouri\\
Columbia, MO, USA\\
krkq35@gmail.com}
}


%


\maketitle

\begin{abstract}
Medical image segmentation has become an essential technique in clinical and research-oriented applications. Because manual segmentation methods are tedious, and fully automatic segmentation lacks the flexibility of human intervention or correction, semi-automatic methods have become the preferred type of medical image segmentation. We present a hybrid, semi-automatic segmentation method in 3D that integrates both region-based and boundary-based procedures. Our method differs from previous hybrid methods in that we perform region-based and boundary-based approaches separately, which allows for more efficient segmentation. A region-based technique is used to generate an initial seed contour that roughly represents the boundary of a target brain structure, alleviating the local minima problem in the subsequent model deformation phase. The contour is deformed under a unique force equation independent of image edges.  Experiments on MRI data show that this method can achieve high accuracy and efficiency primarily due to the unique seed initialization technique.  
\end{abstract}

\begin{IEEEkeywords}
segmentation; visualization; validation; MRI

\end{IEEEkeywords}

%
\IEEEpeerreviewmaketitle

\section{Introduction}
In recent years, medical image segmentation has become a standard technique for visualizing structures of the human brain as well as performing various types of volumetric and shape comparisons among these structures. Since the introduction of medical image segmentation, many methods have been implemented for brain structure segmentation from magnetic resonance imaging (MRI). These methods can be categorized into manual, semi-automatic, and fully automatic methods. Manual segmentation is tedious, requires training and much attention to detail, and the results are not reproducible. On the other hand, fully automatic methods require no training and are completely reproducible for the same data, but these methods do not allow for human intervention or manipulation and severely limit the autonomy of the one performing the segmentation. These issues have caused semi-automatic methods to become the preferred type of medical image segmentation~\cite{medSemiSeg}.

Semi-automatic segmentation can also be divided into several categories, but the two primary classifications for medical image segmentation include region-based (region growing, region merging)~\cite{hybrid4} and boundary-based (snake and balloon)~\cite{met9, met16, qing24} techniques. Region-based methods provide quick segmentation results by assigning membership to voxels according to homogeneity statistics, but the inhomogeneity among MRI voxel intensities can result in inaccurate segmentation (i.e. holes and irregular boundaries)~\cite{hybrid4}. 

Boundary-based methods attempt to align an initial deformable boundary with the object boundary by minimizing an energy functional which quantifies the gradient features near the boundary. This technique works well for images with little noise but these methods are also generally unreliable because image noise and low contrast edges between brain structures can result in false or non-existent boundaries causing under- or over-segmentation ~\cite{met9, met16}. For boundary-based methods, defining the initial geometry prior to deformation (i.e. the seed) is also a critical issue that has yet to be resolved. Without a proper method for seed contour initialization, the seed will deform to local rather than global minima in most circumstances due to image noise. ~\cite{met5, met10} provide evidence for these advantages and drawbacks.  These unrefined seeding methods increase the time required to complete each segmentation.

The most effective way to overcome the issues with region-based and boundary-based methods is to utilize both methods in parallel.~\cite{hybrid2, hybrid3} used region-based information to drive the explicit deformable models in their techniques, while~\cite{hybrid1, hybrid5, hybrid6, qing22, qing21} have addressed these issues by interlacing region-based and boundary-based methods into a united, iterative segmentation process.  The efficacy of these types algorithms exceed that of region-based or boundary-based methods independently, but the most notable disadvantages of these methods are that they limited to slice-by-slice (2D) segmentation and have relatively low efficiency compared with other segmentation algorithms. Also, texture has been used as a criteria to drive a hybrid deformable model as in~\cite{metamorph}.  However, this method also lacks an efficient seed initialization technique.

\section{Overview}
In this paper, we present a hybrid semi-automatic segmentation method in 3D that integrates both region-based and boundary-based procedures.  The goal of this algorithm is to achieve robust, efficient, and reproducible results while avoiding the downfalls of region-based and boundary-based procedures methods individually. Unlike previous methods, we use a clustering technique to initialize the deformable model and then deform the model with a unique force equation independent of image edges. Separating the two methods allows for more efficient segmentation in comparison with previous hybrid procedures. By initializing the model with a region-based technique, we are able to avoid most, if not all, local minima in which the model may fall into during deformation.  The speed of segmentation is also substantially improved because of this near-boundary seeding method.  We utilize an implicit deformable model to alleviate the problem of topology changes, such as self-intersection~\cite{levelsetFedkiw}.  The model is driven by the PDE similar to the one derived in~\cite{contourWOEdges}, which allows the evolution of the model to be less affected by image noise and low-contrast edges.

\section{Methods}
The algorithm consists primarily of two independent phases: seed initialization and implicit deformation.  To initialize the seed, a user must first specify a voxel inside the target structure in the MRI.  Next, the MRI is clustered (using K-Means) based on voxel intensity to obtain a general outline of the structure.  We refine this seed by performing a mixture of mathematical morphology combined with a connected components search. Specifically, the seed obtained from clustering is first eroded several times.  Once erosion is complete, a connected components search originates from the point the user has previously specified, and the voxels that are not found during this search (i.e. not connected to the primary cluster) are removed from seed.  This ensures that the seed is now a connected structure, like those in the MRI we wish to segment.  A matching number of dilation steps are performed on the seed to ensure that the seedÕs volume has not diminished to an insufficient size.  These steps remove pieces of the cluster that may not be a part of the target structure.
	
\begin{figure}[t]
\begin{center}
\includegraphics[width=0.9\linewidth]{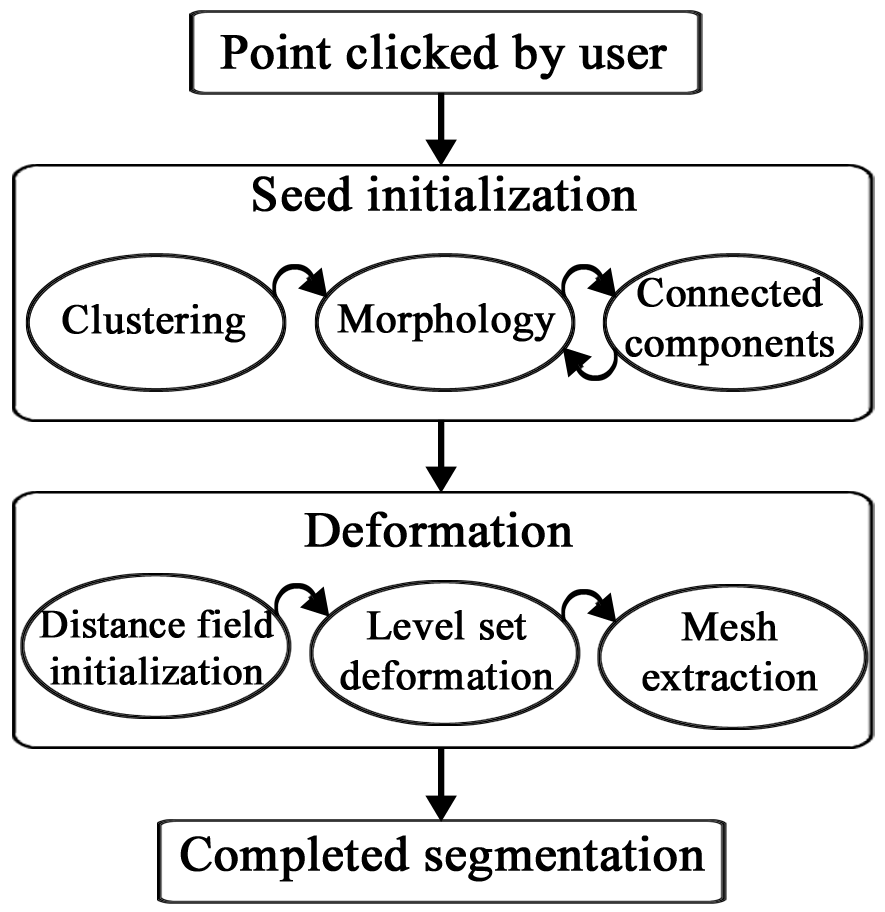}
\end{center}
\caption{Progression through each phase of the semi-automatic segmentation algorithm.  First, a point is specified by the user.  Next, the seed is created by clustering the 3D image, eroding the seed cluster (morphology), ensuring connectedness, and then dilating the seed cluster (morphology).  Then, the seed is passed to the deformation phase where a signed distance field is initialized and deformed, and an explicit mesh is then extracted from the deformed distance field. This mesh represents the completed brain structure segmentation.}
\label{workflowFig}
\end{figure}

To remove artifacts from the seed and complete the segmentation, the seed is deformed based on an implicit level set PDE.  Before deformation, the fast sweeping method for distance field initialization is used to create an implicit representation of the seed.  The distance field is a discrete scalar function that determines the distance from a voxel to the nearest point on the boundary of the implicit surface. Once a distance field has been created, the field is deformed according to a PDE that does not rely heavily on image edges, but rather takes into account surrounding voxel information.  This type of PDE is used to ensure accurate segmentations on structures that have low-contrast edges, such as the hippocampus. A narrow band algorithm for level set evolution is also employed during the deformation in order to increase the efficiency of the segmentation.  Once the equilibrium of the PDE is reached, the segmentation is complete and an implicit representation of the structure is obtained.  We use a standard marching cubes algorithm to obtain an explicit mesh representation of the object for later shape and volumetric comparison. Figure~\ref{workflowFig} diagrams the main phases of the algorithm as well as the intermediate stages within each phase.

	\subsection{Seed Generation}
One of the most important process of the algorithm is the initialization process in which a suitable seed is determined for the deformable model. For the purposes of this paper, the seed is defined as the initial geometry prior to deformation (i.e. $t=0$). The seed created during this initial phase provides a coarse representation of the target shape. We are able to create this seed using a simple clustering technique as well as several iterations of connected components and mathematical morphology.

Let $I$ be a mapping such that $I:\mathrm{\Omega} \rightarrow[0,1]$ where $\mathrm{\Omega}$ is defined by the dimensions of the 3D MRI.  In order to ensure that $I$ is valued in the interval $[0,1]$, we normalize the MRI data by setting all voxel intensities below the $2^{nd}$ percentile to $0$, all intensities above the $98^{th}$ percentile to $1$, and linearly interpolating the values in between so that all voxel intensities fall between $0$ and $1$.   To initiate the algorithm, a user must specify a point inside a target structure ($\mathbf{x}_0\in\mathrm{\Omega}$).

The next step is to cluster $\mathrm{\Omega}$ using the k-means clustering algorithm as found in ~\cite{kmeans}. This is done by minimizing the error term $\mathcal{E}$:
\begin{equation}
  \mathcal{E} = \sum_{j=1}^k \sum_{\mathbf{x}\in{C_j}}|I(\mathbf{x})-I(\mathbf{\tilde{x}}_j)|
\end{equation}
\noindent
where $k$ is the number of clusters to be used, each $C_j$ is a subset of $\mathrm{\Omega}$ and disjoint from one another such that $\bigcup_{j=1}^k C_j = \mathrm{\Omega}$, and $\mathbf{\tilde{x}}_j$ is the average intensity of all voxel intensities in the cluster $C_j$.  Either a mean shift algorithm can be applied to $I$ to determine $k$, or a user can specify this value.

Clustering separates the domain, $\mathrm{\Omega}$ into $k$ clusters.  The most important cluster is the one which contains $\mathbf{x}_0$ since this cluster provides an initial estimate of the target shape. Thus, the other clusters can be disregarded.  Assume that $\mathbf{x}_0 \in{C_m}$ where $m\in{\{1,\ldots,k\}}$.  Then, let ${\mathcal{I}}=C_m$ and ${\mathcal{O}} = \mathrm{\Omega}\setminus C_m$. We use $\mathcal{I}$ to denote the voxels \textit{inside} of the deformable model, and $\mathcal{O}$ to denote the voxels \textit{outside} of the model.  This notation is important when discussing the future level-set deformation of the model.  The next steps will further refine the seed ($\mathcal{I}$) by ensuring it's connectedness and removing voxels from the seed which are not strongly affiliated with the target structure.

Voxels which are not affiliated with the target structure will in most cases be in or near the boundary of ${\mathcal{I}}$, denoted $\partial {\mathcal{I}}$. To remove these voxels, ${\mathcal{I}}$ is eroded with a mathematical morphology operation a certain number of iterations (dependent on the size of $\mathrm{\Omega}$). This essentially removes $\partial {\mathcal{I}}$ from ${\mathcal{I}}$ at each iteration, and then recalculates ${\mathcal{I}}$ after each step of erosion. In some cases, essential voxels of the target structure are removed from $\mathcal{I}$, but the deformation stage (see section 3.2) is used to overcome these seeding artifacts.

As with most mathematical morphology operations, $\mathcal{I}$ should then be dilated the same number of steps that it was eroded.  However, in our implementation, this step is proceeded by a connected components step to ensure the connectedness of $\mathcal{I}$ because there is no guarantee, and little possibility, that ${\mathcal{I}}$ is connected. Since most structures in the brain are connected, we make an  assumption that our seed should be connected as well to produce more accurate segmentations. To alleviate this possible unconnectedness, a simple connected components algorithm is implemented originating from the point $\mathbf{x}_0$.  During the connected components search, only voxels in $\mathcal{I}$ are available to be searched.  $\mathcal{I}$ is recalculated based on all voxels that are visited during the search, and the outside set is also recalculated with the equation ${\mathcal{O}} = \mathrm{\Omega}\setminus \mathcal{I}$.

\begin{figure}[t]
\begin{center}
\includegraphics[width=0.9\linewidth]{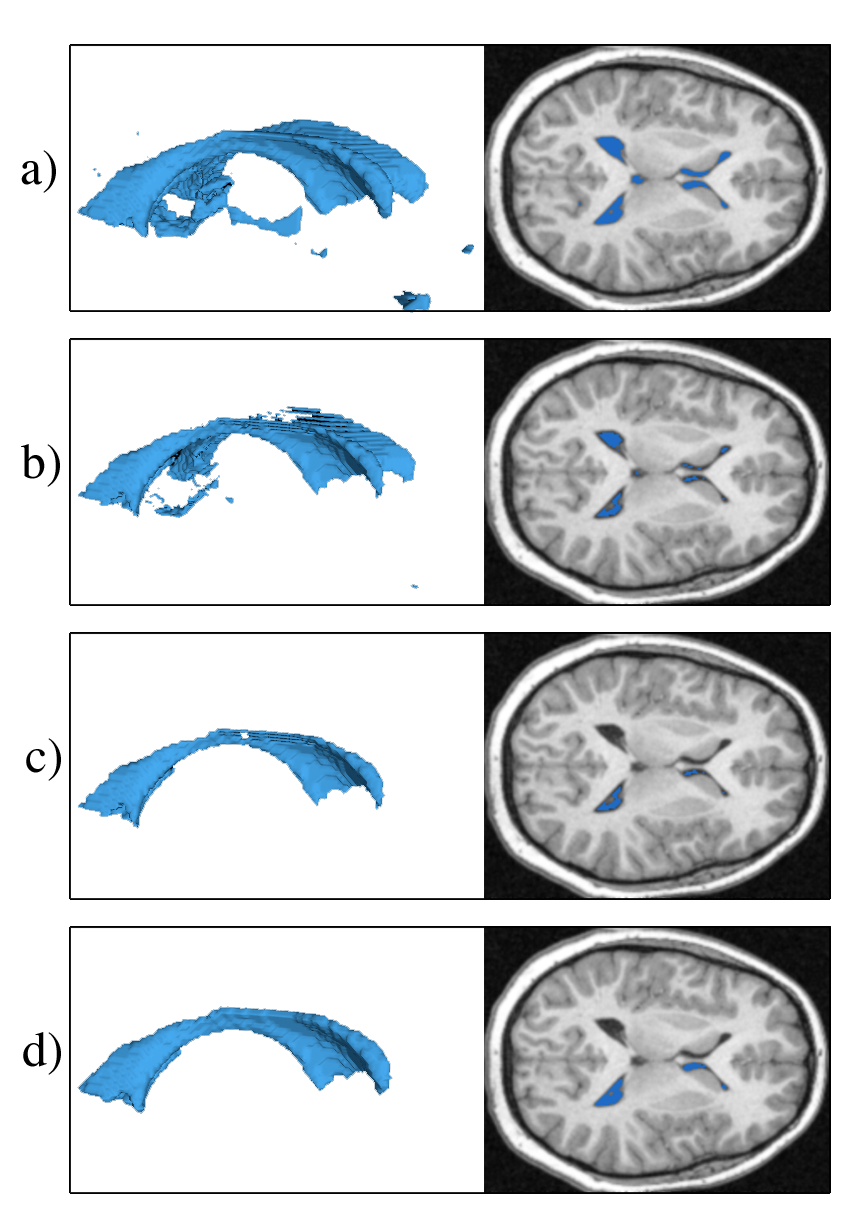}
\end{center}
\caption{Demonstration of the clustering, erosion, connected components, and dilation stages of the seed generation phase during segmentation of a right lateral ventricle (chronologically, from top to bottom). The 3D seed (left) and a 2D slice from the MRI (right) are rendered corresponding to the current stage. On the 2D image slice, pixels are rendered as blue if they belong to the seed. a) seed after K-means clustering; b) seed after erosion; c) seed after connected components search; and d) seed after dilation.}
\label{seedingFig}
\end{figure}

Finally, the set ${\mathcal{I}}$ is dilated the same number of steps that it was eroded, recalculating ${\mathcal{I}}$ and ${\mathcal{O}}$ dynamically.  These seed creation steps can be done very efficiently, and allow for the development of a seed that is roughly equivalent to the target structure. Figure~\ref{seedingFig} shows the progression of the seed initialization phase from start to finish.

	\subsection{Deformation}

\begin{figure}[t]
\begin{center}
\includegraphics[width=0.9\linewidth]{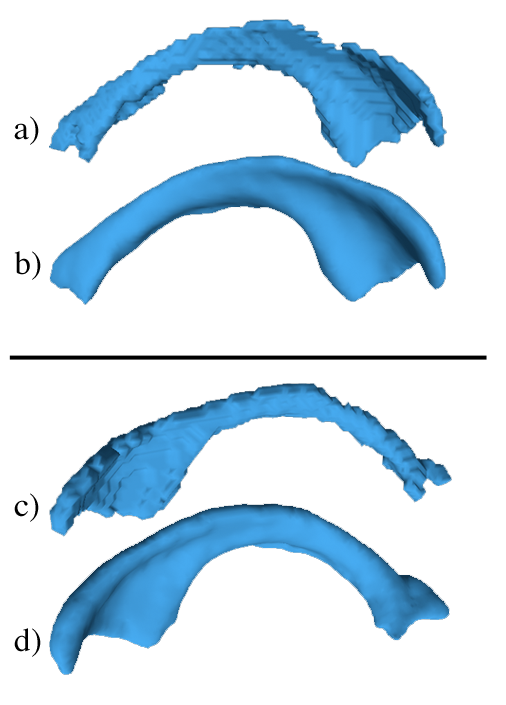}
\end{center}
\caption{Seed contours rendered before and after level-set deformation.  a) right lateral ventricle seed prior to deformation; b) completed segmentation after deforming the generated seed (a).  c) left lateral ventricle seed prior to deformation; d) completed segmentation after deforming the generated seed (c). Notice that the deformation phase removes holes, and non-smoothness from the initial seed.}
\label{deformationFig}
\end{figure}

Some seeds will be coarse and also contain various artifacts or imperfections due to image noise. To increase the accuracy of segmentation, a subsequent deformation phase is applied to the initial seed.  Figure~\ref{deformationFig} illustrates the necessity and benefits of this subsequent deformation phase.

To begin deformation, the seed, $\mathcal{I}$ is transformed into an implicit distance field first and will deform based on a level-set PDE. A signed distance function, $\phi:\mathrm{\Omega}\rightarrow \mathbb{R}$ is initialized such that $\phi(\mathbf{x})$ is the signed euclidean distance (positive on the set $\mathcal{O}$ and negative on the set $\mathcal{I}$) from $\mathbf{x}$ to the closest voxel in $\partial \mathcal{I}$. 
	
We have found that the most efficient method to perform this distance field initialization is to use a 3D fast sweeping method as described in in~\cite{fastSweeping}. The fast sweeping method works by first initializing $\phi$ to be very large at all points in its domain.  Then, directly compute $\phi$ for all voxels that are in or adjacent to $\partial \mathcal{I}$ with the following equation:
\begin{equation} 
\phi(\bf{x}) = \left\{ \begin{array}{l} 
\displaystyle{-\min_{\bf{y}\in{\mathcal{\partial \mathcal{I}}}}}\|\bf{x}-\bf{y}\|	\; 	\mathrm{if}\, \bf{x}\in{\mathcal{I}} \\ 
\displaystyle{\min_{\bf{y}\in{\mathcal{\partial \mathcal{I}}}}}\|\bf{x}-\bf{y}\|	\; \mathrm{if}\, \bf{x}\in{\mathcal{O}} 
\end{array}\right. 
\end{equation}
\noindent
The rest of $\phi$'s domain is computed by propagating an approximation of the actual distance using elements which have previously been computed.  To achieve this, the domain is "swept" a total of eight times (starting from a new corner of the 3D rectangle and ending at the opposite corner for each sweep) using the below method of computation to solve for the values of $\phi$. In three dimensions, this calculation is done by solving the following equation at each grid point, assuming a uniform grid size of one:
\begin{equation} 
[(x-a_1)^+]^2 + [(x-a_2)^+]^2 + [(x-a_3)^+]^2 = 1
\end{equation}
\noindent
Where $(m)^+$ is the maximum between $m$ and 0, $x$ is the value of $\phi$ at the point in question, and $a_1, a_2, a_3$ are the minimum surrounding values in each direction of the point such that $a_1 \leq a_2 \leq a_3$. For a detailed explanation of the fast sweeping algorithm in $n$ dimensions, refer to~\cite{fastSweeping}. To make $\phi$ a true signed distance field, it must be ensured that  $\phi$ is negative when acting on any voxels in $\mathcal{I}$ and positive for all values in $\mathcal{O}$.
	
$\phi$ is then deformed based on a PDE similar to those described in~\cite{contourWOEdges,fastHybrid}.  The PDEs in these works focus heavily on using information other than edge detection functions to drive the level-set deformation.  The idea behind the deformation is to introduce an artificial time variable $t$ and to update the level-set function $\phi$  as time elapses.  We update $\phi$ by numerically solving the following PDE at each time step:
\begin{equation}
  \frac{\partial{\phi}}{\partial{t}} = \alpha \nabla \cdot \left( \frac{\nabla\phi}{\|\nabla\phi\|}\right) -\beta -\gamma_1(I - \tilde{\mathcal{I}})^2 + \gamma_2(I - \tilde{\mathcal{O}})^2
\end{equation}
\noindent
where $\tilde{\mathcal{I}}$ and $\tilde{\mathcal{O}}$ are the arithmetic means of the intensities of all voxels in their respective sets, and $\alpha$, $\beta$, $\gamma_1$, and $\gamma_2$ are coefficients.   More specifically, $\alpha$ determines the weight to be associated with the curvature term, $\beta$ is an external force to be applied to $\phi$, and $\gamma_1$ and $\gamma_2$ are weights for the distance functions $(I - \tilde{\mathcal{I}})^2$ and $(I - \tilde{\mathcal{O}})^2$ respectively.  It is important to note that the sets $\mathcal{I}$ and $\mathcal{O}$ are updated at each iteration to match the way they had been previously defined with respect to $\phi$, i.e. $\mathcal{I}=\{\bf{x}\in\mathrm{\Omega}: \phi({\bf x}) \leq \rm{0}\}$ and $\mathcal{O}=\{\bf{x}\in\mathrm{\Omega}: \phi({\bf x}) > \rm{0}\}$.  Thus, $\tilde{\mathcal{I}}$ and $\tilde{\mathcal{O}}$ will need to be recalculated at each iteration as well.

For the purpose of efficiency, we also implement a narrow band algorithm so that only values of the distance field that are within a certain threshold are updated.  For example, while updating the distance field, we only solve the aforementioned PDE near the voxel points where $\phi=0$. A similar approach and more detailed implementation can be found in~\cite{narrowBand}. 

Once the deformation is complete, the signed distance function $\phi$ now provides us with an implicit representation of the target shape; however, an explicit mesh must be extracted from $\phi$ for further shape analysis and shape comparison.  For our implementation, this is done with a standard marching cubes algorithm as discussed in~\cite{marchingCubes}.  This algorithm takes a three dimensional signed distance function as input and returns a mesh of triangles and vertices that represents a discrete approximation of the zeroth level-set of the distance function.

\section{Results and Validation}
\begin{figure}[t]
\begin{center}
\includegraphics[width=0.9\linewidth]{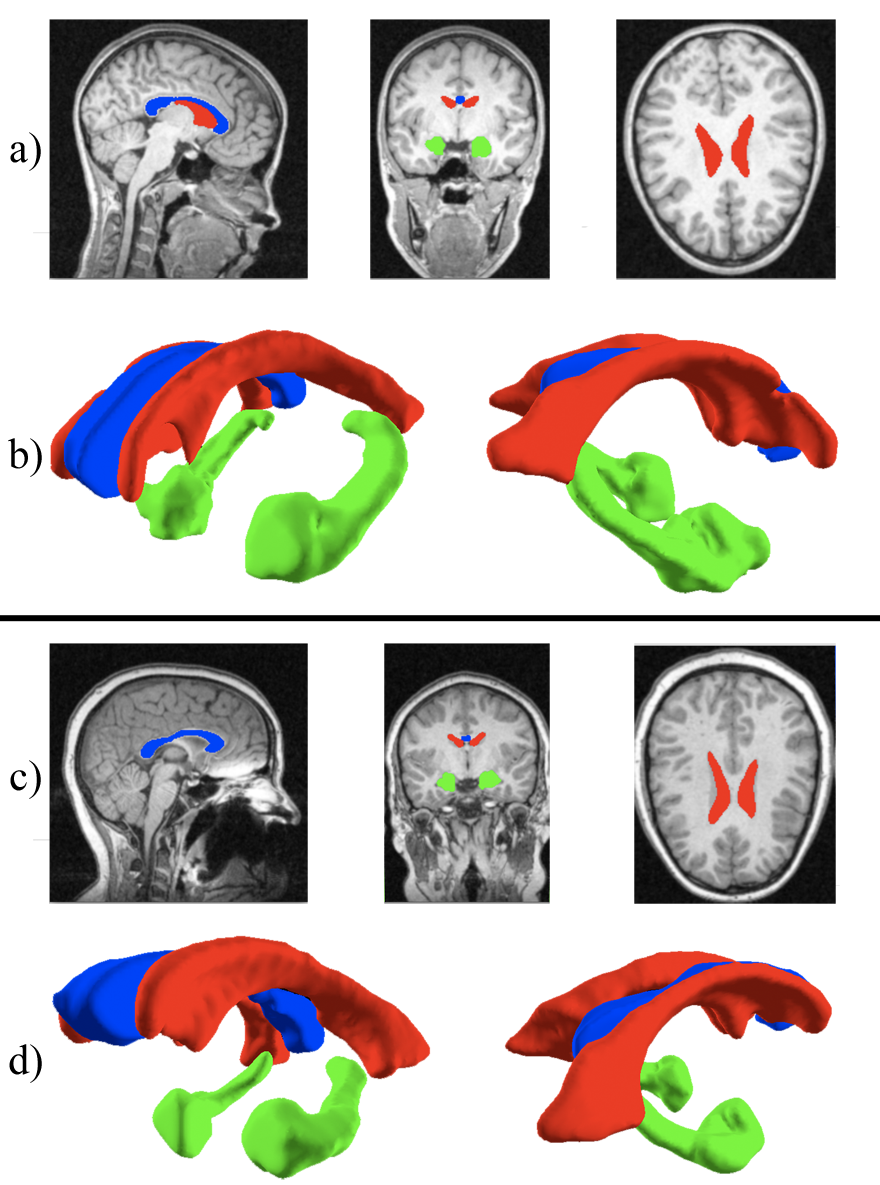}
\end{center}
\caption{Completed segmentations from two different MRI (top and bottom) using the technique presented in this paper, rendered in different orientations and perspectives. a) and c) display results superimposed on 2D slices in the sagittal (left), coronal (middle), and axial (right) planes. b) and d) show the segmentations rendered as 3D objects from two different perspectives.  Blue corresponds to the corpus callosum, red corresponds to the left and right lateral ventricles, and green corresponds to the left and right hippocampi for both MRI (best viewed electronically).}
\label{conclusionFig}
\end{figure}

This method has been used to segment various brain structures from actual patient MRI as well as test data.  We have successfully segmented the corpus callosum, lateral ventricles, hippocampi, and thalami from these data sets with few failures.   Figure~\ref{conclusionFig} shows several examples of completed segmentations from several different orientations and perspectives.

To test the validity and accuracy of the segmentations, we have compared segmented structures using our algorithm to manual (ground truth) segmentations of the structure from the same MRI.  Only real patient MRI were used in these experiments to verify that our technique is applicable in clinical research settings.  For the comparisons, we calculate volumetric comparison statistics (dice similarity and overlap coefficients) between the two segmentations.  The range of these statistics lie between 0 and 1, with 1 indicating a perfect agreement between our segmentation and the manual segmentation. We tested the corpus callosum, left and right ventricles, and left and right hippocampi segmentations using our algorithm against corresponding manual segmentation of these structures performed by a trained expert to obtain the dice similarity and overlap coefficients.  Overall, our method was able to closely reproduce the accuracy of the manual segmentations. Table~\ref{table1} provides a summary of the results.

\begin{table}
\begin{center}
\caption{Comparisons of the semi-automatic segmentation algorithm presented in this paper to manual (ground truth) segmentations.  N denotes the number of segmentations used for each structure.  $\mu_D$ is the arithmetic mean of the dice similarity calculation for each structure, and $\sigma_D$ is the standard deviation for these calculations.  Similarly, $\mu_O$ is the arithmetic mean of the overlap coefficient calculation for each structure, and $\sigma_O$ is the standard deviation for these calculations. For both the dice similarity and overlap coefficient means, each value ranges from 0 to 1, with larger values indicating better results.}
\begin{tabular*}{0.45\textwidth}{@{\extracolsep{\fill}} |c|ccccc|}
\hline
Structure & N & $ \mu_D$ & $\sigma_D$ & $\mu_O$ & $\sigma_O$ \\ \hline
Corpus & \multirow{2}{*}{16} & \multirow{2}{*}{0.79} & \multirow{2}{*}{0.05} & \multirow{2}{*}{0.79} & \multirow{2}{*}{0.03} \\ 
callosum & & & & & \\ \hline
Lateral ventricle& \multirow{2}{*}{8} & \multirow{2}{*}{0.81} & \multirow{2}{*}{0.04} & \multirow{2}{*}{0.73} & \multirow{2}{*}{0.05} \\ 
(left) & & & & & \\ \hline
Lateral ventricle& \multirow{2}{*}{8} & \multirow{2}{*}{0.78} & \multirow{2}{*}{0.06} & \multirow{2}{*}{0.81} & \multirow{2}{*}{0.03} \\ 
(right) & & & & & \\ \hline
Hippocampus & \multirow{2}{*}{8} & \multirow{2}{*}{0.69} & \multirow{2}{*}{0.08} & \multirow{2}{*}{0.66} & \multirow{2}{*}{0.06} \\ 
(left) & & & & & \\ \hline
Hippocampus & \multirow{2}{*}{8} & \multirow{2}{*}{0.70} & \multirow{2}{*}{0.07} & \multirow{2}{*}{0.65} &  \multirow{2}{*}{0.05} \\ 
(right) & & & & & \\ \hline
\end{tabular*}
\end{center}
\label{table1}
\end{table}

We have also compared the efficiency of our segmentation method to a boundary-based method using a generic seed (a sphere centered at the user-specified point).  The only difference between these two methods was the way in which the seed was created. The results are shown in Table~\ref{table2}, where Method A uses with our seeding method for segmentation, and Method B uses the aforementioned generic seeding method for segmentation. The results show a large increase in efficiency when using the seeding technique described in this paper, with a speed-up of nearly 300 to 500 percent based on the structure.  Combined with the accuracy our algorithm attains, this efficiency comes at little to no cost to the overall segmentation results.

\begin{table}
\begin{center}
\caption{Comparisons of the efficiency of segmentation using differing seeding methods.  Method A was seeded with the technique described in this paper; Method B was seeded by initializing a sphere centered at a user-specified point.  All other aspects of the segmentation remained constant.  The amount of time each segmentation took is measured in seconds.}
\begin{tabular*}{0.45\textwidth}{@{\extracolsep{\fill}} |c|cc|}
\hline
Structure & Method A & Method B \\ \hline
Corpus callosum & 33s & 102s \\ \hline
Lateral ventricle (left) & 54s & 263s \\ \hline
Lateral ventricle (right) & 58s & 240s \\ \hline
Hippocampus (left) & 47s &  142s \\ \hline
Hippocampus (right) & 46s & 148s \\ \hline
\end{tabular*}
\end{center}
\label{table2}
\end{table}

\section{Conclusion and Future Work}
Differing from previous hybrid segmentation methods, this paper presents a technique for seeding that can improve the accuracy and efficiency of current deformable model semi-automatic segmentation methods.  This initialization method is novel and can substantially increase the segmentation time of modern deformable geometry segmentation, both explicit and implicit. We have also introduced a new method for combining region-based and boundary-based segmentation methods, which is significantly different than the previous hybrid methods. This semi-automatic segmentation algorithm maintains accuracy while significantly improving the speed of modern boundary-based segmentation; the validation results attest to this. The method works with a number of structures in the brain and is not limited to a specific structure. Also, the algorithm is designed to segment 3D structures outright, as opposed to performing 2D slice-by-slice segmentation and stitching the resulting contours together to obtain a 3D shape.

While our algorithm has been able provide consistent results among the data sets we have tested, there have been a few failed segmentations.  This was primarily caused by excessive noise in the MRI, or due to low image resolution.  In the future, we hope to address these issues by perfecting the seeding and deformation phases, or perhaps adding several MRI preprocessing stages to our algorithm.  We would also like to increase the robustness of our algorithm to allow for segmentation of several other complex structures, such as the cerebellum, by incorporating a texture parameter into the PDE in which our deformable model is governed by, such as in~\cite{metamorph}.

\section*{Acknowledgment}
This work is supported in part by a NIH pre-doctoral training grant for Clinical Biodetectives, the Department of Defense Autism Concept Award, and the NARSAD Foundation Young Investigator Award.

{\small
\bibliographystyle{IEEEtran}
\bibliography{SegmentationBib}
}

\end{document}